\documentclass[letterpaper, 10 pt, conference]{ieeeconf}  

\IEEEoverridecommandlockouts                              

\usepackage{graphicx}		
\usepackage{amsmath}

\usepackage{booktabs}
\usepackage{multirow}
\usepackage{amssymb}	
\usepackage[detect-all,per-mode=symbol]{siunitx}

\usepackage{standalone}
\usepackage{pgfplots}
\pgfplotsset{compat=newest}
\pgfplotsset{every axis legend/.append style={legend cell align=left}}
\usepgfplotslibrary{groupplots}        
\usepgfplotslibrary{polar}            

\usepackage[keeplastbox]{flushend}

\usepackage{commath}
\usepackage[capitalize]{cleveref}		

\newcommand{\domain}{X}
\newcommand{\kk}{\mathbf{k}}
\newcommand{\phik}{\phi_\kk}
\newcommand{\Fk}{F_\kk}
\newcommand{\xa}{x_a}
\newcommand{\xb}{x_b}
\newcommand{\ck}{c_\kk}

\newcommand{\cka}{c_\kk^a}
\newcommand{\ckb}{c_\kk^b}

\newcommand{\ca}{c^a}
\newcommand{\cb}{c^b}

\newcommand{\Ta}{T_a}
\newcommand{\Tb}{T_b}

\newcommand{\point}{\mathbf{x}}

\newcommand{\dt}{\dif{t}}

\newcommand{\dpoint}{\dif{\point}}

\newcommand{\Lambdak}{\Lambda_\kk}
\newcommand{\I}{\mathcal{I}}

\title{\LARGE \bf
On the Optimality of Ergodic Trajectories \\for Information Gathering Tasks
}

\author{Louis Dressel and Mykel J. Kochenderfer
\thanks{This work was supported NSF grant DGE-114747.}
\thanks{The authors are with the Aeronautics and Astronautics department at Stanford University, Stanford, CA, 94305 USA. Email:
		{\tt\small \{dressel,mykel\}@stanford.edu}}%
}

\begin{document}

\maketitle
\thispagestyle{empty}
\pagestyle{empty}

\begin{abstract}
Recently, ergodic control has been suggested as a means to guide mobile sensors for information gathering tasks.
In ergodic control, a mobile sensor follows a trajectory that is ergodic with respect to some information density distribution.
A trajectory is ergodic if time spent in a state space region is proportional to the information density of the region.
Although ergodic control has shown promising experimental results, there is little understanding of why it works or when it is optimal.
In this paper, we study a problem class under which optimal information gathering trajectories are ergodic.
This class relies on a submodularity assumption for repeated measurements from the same state.
It is assumed that information available in a region decays linearly with time spent there.
This assumption informs selection of the horizon used in ergodic trajectory generation.
We support our claims with a set of experiments that demonstrate the link between ergodicity, optimal information gathering, and submodularity.

\end{abstract}

\section{INTRODUCTION}
In information gathering tasks, a mobile sensing agent plans a trajectory that maximizes the information gathered about the environment.
The gathered information is typically measured as a reduction in uncertainty.
Target localization is a type of information gathering task where the agent makes observations while searching for a target.
The agent maintains a belief, which is a probability distribution over possible target locations.
This belief is typically updated using Bayes' rule and a measurement model.
The agent's goal is to gather information about the target's location, which reduces uncertainty in the target estimate (measured by belief entropy or variance).
Information gathering tasks appear in many real-world problems.
Examples include localizing GPS jammers~\cite{Perkins2015}, radio-tagged wildlife~\cite{cliff2015}, or disaster victims~\cite{Hoffmann2006}.

Unfortunately, information gathering tasks are computationally challenging.
Multi-step planning problems can be cast as partially observable Markov decision processes (POMDPs), which are computationally intractable to solve exactly~\cite{papadimitriou1987,madani2003}.
One approach is to approximate the belief as a Gaussian and linearize the dynamic and measurement models.
These approximations allow trajectories to be evaluated quickly in a model predictive control (MPC) framework~\cite{Leung2005,Leung2006}.
However, these approximations are not always appropriate for nonlinear systems with non-Gaussian beliefs.
Other approaches avoid linear-Gaussian assumptions but optimize for a single time step rather than an entire trajectory.
Although these greedy, information-theoretic approaches are suboptimal in general, they are computationally efficient and often used in practice~\cite{Hoffmann2006}.

Ergodic control has recently been proposed for designing trajectories for mobile sensors~\cite{silverman2013,miller2016,millerthesis2}.
This framework can be applied to general, nonlinear systems and has outperformed greedy methods in some experiments~\cite{miller2015,miller2016}.
Ergodic control is built on the notion of trajectory ergodicity.
A trajectory is ergodic with respect to some distribution if time spent in a state space region is proportional to the distribution's density in that region.
When using ergodic control for information gathering, the distribution used is an expected information density, which is a measure of information at a point in the sensor's state space.

Although ergodic control has shown promising experimental results, it has only recently been applied to information gathering tasks.
It is not understood \textit{why} ergodic control works well.
Why does it make sense to spend time in a region proportional to its information density, instead of spending all our time in the most dense region?
Selecting the length of an ergodic trajectory is another open research problem~\cite{silverman2013}.

In this paper, we provide some insight into these fundamental questions of ergodic control.
We present a problem class for which the optimal information gathering trajectory is ergodic.
This class assumes measurement submodularity, where successive measurements from a state reduce the information available at that state.
Specifically, the class assumes the rate of decay is linear.
Under this assumption, selection of the ergodic optimization horizon for many systems becomes trivial.
We use simple toy problems to validate these ideas and show the potential suboptimality of ergodic control when the assumptions do not hold.
We generate ergodic trajectories for more complex problems to verify the connection between optimal information gathering, information decay, and ergodic trajectories.

\section{ERGODIC CONTROL}
\label{sec:background}

\begin{figure}
	\centering
	\includegraphics[trim={1.1in .3in 1.1in .5in},clip,width=1.6in]{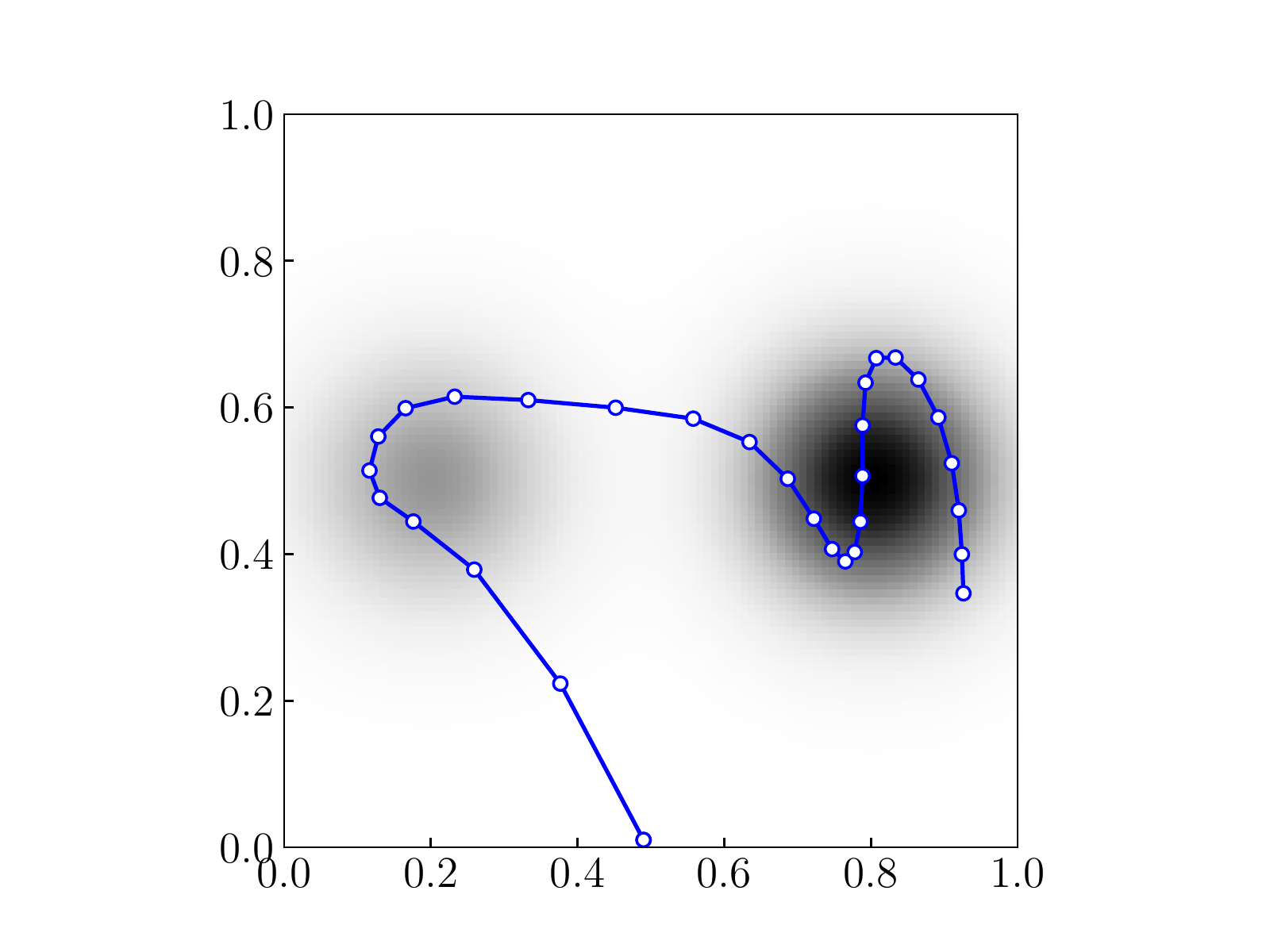}
	\includegraphics[trim={1.1in .3in 1.1in .5in},clip,width=1.6in]{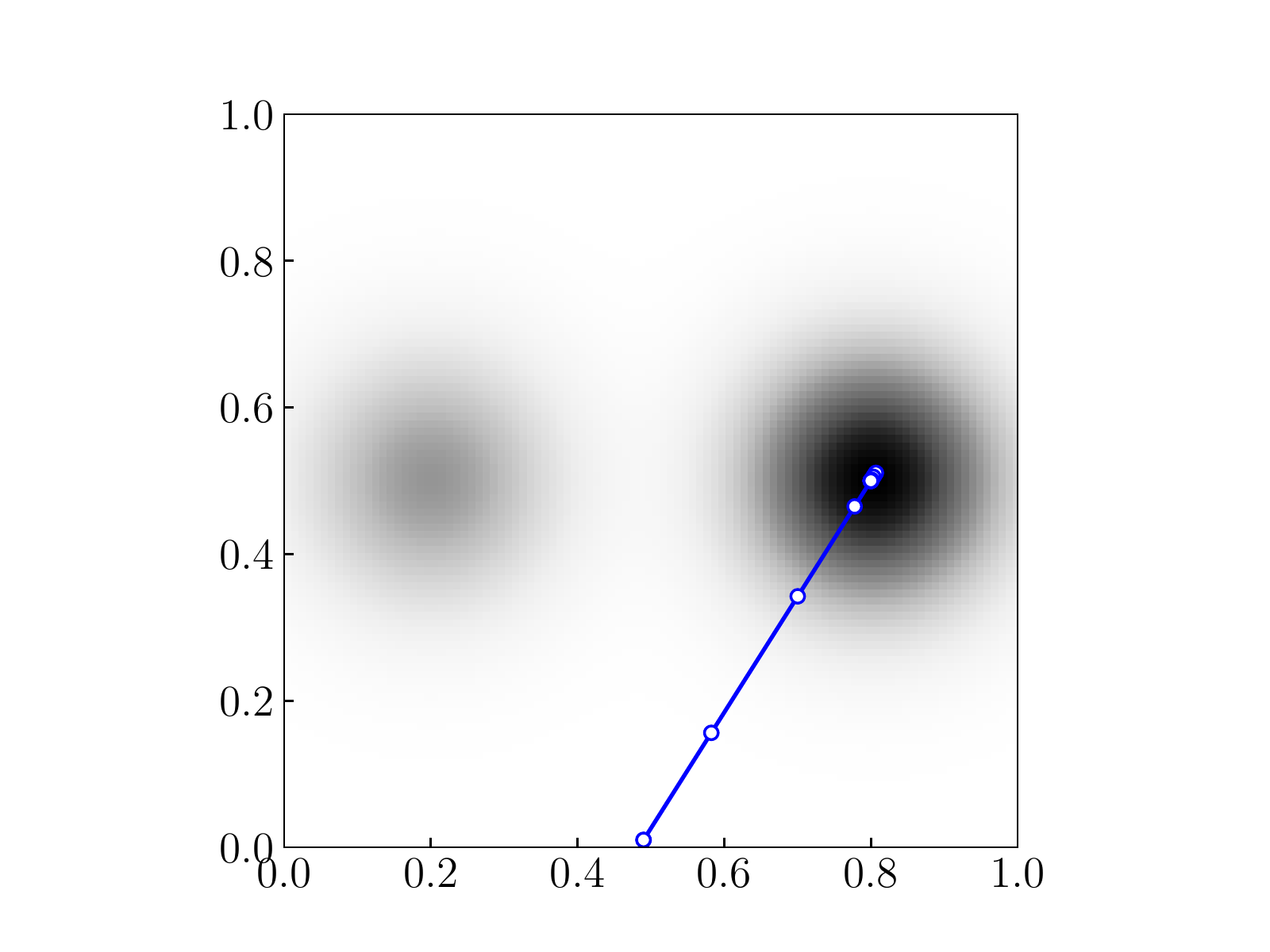}
	\caption{An ergodic trajectory (left) and a trajectory that simply moves to the highest density state (right). Both trajectories start from $(0.5,.01)$.}
	\label{fig:ergovsmax}
\end{figure}

Ergodic control relies on the concept of trajectory ergodicity.
A trajectory is ergodic with respect to a distribution if its time-averaged statistics match the distribution's spatial statistics.
In other words, the time spent in a region is proportional to the distribution's density in the region.
\cref{fig:ergovsmax} compares a trajectory ergodic with a distribution and a trajectory maximizing time spent in high density regions.

We use the following notation.
Consider a domain $\domain\subset\mathbb{R}^s$ and a distribution $\phi:\domain \to \mathbb{R}$ that provides a density $\phi(\point)$ at a state $\point\in\domain$.
A trajectory of horizon $T$ is a function $x:[0,T]\to\domain$.
The state at time $t$ according to trajectory $x$ is denoted $x(t)$.

The time-averaged statistics of a trajectory are a distribution $c$ over the state space, where the density at $\point$ is
\begin{equation}
	c(\point) = \frac{1}{T}\int_0^T \delta(\point - x(t))\dt\text,
\end{equation}
where $\delta$ is the Dirac delta function.
The factor $1/T$ ensures the distribution integrates to 1.
Likewise, $\phi$ must be a valid density that integrates to 1 so that $c$ and $\phi$ can be compared.

The goal in ergodic control is to drive $c$ to equal $\phi$.
This goal is made explicit in an ergodic metric that measures the KL divergence between $c$ and $\phi$~\cite{ayvali2017}.
The KL divergence measures the similarity of two distributions.
A different but widely-used ergodic metric decomposes $c$ and $\phi$ into Fourier coefficients and compares the coefficients to each other~\cite{mathew2011}.
The distribution is decomposed into Fourier coefficients $\phik$:
\begin{equation}
	\phik = \int_X \phi(\point)\Fk(\point)\dpoint,
\end{equation}
where $\Fk$ is a Fourier basis function and $\kk=[k_1,\dots,k_s]$ is a multi-index used to simplify notation; $\phi_{\kk}$ is short for $\phi_{k_1,k_2,\dots,k_s}$. 
Each $k_i$ ranges from 0 to $K$; there are $(K+1)^s$ coefficients in total.
The coefficients $\ck$ of trajectory $x$ are
\begin{equation}
	\ck(x) = \frac{1}{T}\int_0^T \Fk(x(t))\dt\text.
\end{equation}
The ergodic metric $\mathcal{E}$ is a weighted sum of the squared difference between trajectory and distribution coefficients:
\begin{equation}
	\label{eq:metric}
	\mathcal{E}(x) = \sum_\kk \Lambdak \norm{\ck(x) - \phik}^2\text,
\end{equation}
where $\sum_\kk$ is short for $\sum_{k_1=0}^{K}...\sum_{k_s=0}^{K}$ and weights $\Lambdak$ favor low-frequency features.
This metric has been used in feedback laws that drive trajectories toward ergodicity~\cite{mathew2011}.

Strictly speaking, trajectories are only ergodic if $c\to\phi$ as $T\to\infty$~\cite{mathew2011}.
However, we follow recent work and call trajectory $x$ ergodic if $\mathcal{E}(x)$ is small, even for finite horizons.
Projection-based trajectory optimization (PTO) is one way to design ergodic trajectories for a given horizon $T$~\cite{miller2013}.
This method can be used for general nonlinear systems and the resulting ergodic trajectories have been used in information gathering tasks~\cite{silverman2013}.
In these tasks, the distribution $\phi$ is an expected information density (EID) that represents the value of making a measurement from a specific state.
The EID can be generated from information-theoretic concepts such as Fisher information or expected entropy reduction.

An ergodic trajectory is open-loop---a trajectory is designed for an EID, but this distribution changes as measurements are made and the belief is updated.
To take advantage of this updated information, an MPC framework can be used~\cite{silverman2013}.
First, an ergodic trajectory is generated for planning horizon $T$.
Then some or all of that trajectory is executed, and measurements are collected.
The belief and EID are updated, and a new ergodic trajectory is generated for planning horizon $T$.
This approach leverages the ability to plan entire trajectories while incorporating updated information.
Because ergodic trajectory generation can be computationally expensive, the execution horizon is often as large as the planning horizon in practice~\cite{silverman2013,millerthesis2,miller2016}.

It has been claimed that ergodic control effectively balances exploration and exploitation of information---more time is spent at information dense regions, but less dense regions are also explored~\cite{miller2016}.
Empirically, ergodic control seems like a viable choice for localization tasks.
When compared to greedy, information-theoretic methods, ergodic control has slightly underperformed when noise is low, but significantly outperformed in environments with significant unmodeled noise~\cite{millerthesis2}.
At extraordinarily high levels of noise, ergodic control has underperformed uniform sweeps of the environment.
When noise is so high as to render the model useless, it is reasonable to cover the space uniformly.
Although it has slightly underperformed greedy and uniform methods when noise is very low or very high, ergodic control generally performs well across noise regimes.
The ability to adapt to concentrated information (low noise) or diffuse information (high noise) is a benefit of ergodic control.


\section{OPTIMALITY AND SUBMODULARITY}
\label{sec:submodularity}

An optimal information gathering trajectory maximizes $\I(x)$, the information gathered by trajectory $x$, while adhering to dynamic or time constraints.
On the surface, it is not clear why an ergodic trajectory would maximize $\I(x)$.
If $\phi(\point)$ represents the information at point $\point$, directly maximizing $\int_0^T \phi(x(t))\dt$ seems reasonable.
This strategy would direct the sensor to the point with highest information density, instead of distributing measurements ergodically.
To justify ergodic behavior, we look to submodularity.

\subsection{Submodularity}
In the context of information gathering, measurement submodularity refers to the notion that repeated measurements from a given location are successively less informative~\cite{miller2016}.
Formally, we say this submodularity is present if $\I(\xa + \xb) \leq \I(\xa) + \I(\xb)$, where $\xa + \xb$ is the concatenation of trajectories $\xa$ and $\xb$~\cite{hollinger2014}.

Submodularity is present in many information gathering tasks and must be accounted for to prevent solely and repeatedly sampling the maximally dense point~\cite{miller2016}.
If the sensor only samples this point, and the information there becomes depleted, the total information gathered along the trajectory might be low.
In one information gathering example with a discrete number of states, the planner assumes a state's information is depleted after a single measurement, preventing sensors from staying at the information maxima~\cite{hollinger2014}.
Another way to handle submodularity is to plan for a single step.
In greedy, one-step trajectory planners, the belief and EID can be updated after each measurement, thereby incorporating submodularity and preventing a sensor from sampling a point with depleted information.
By only planning for the next measurement location, the planner can ignore submodularity induced by an entire trajectory.
However, when planning an entire trajectory for an initial EID, we need something to handle the submodularity.

In this context, it seems that ergodic control might be one way to incorporate submodularity into trajectory generation.
In ergodic control, a trajectory is generated for an initial EID, which becomes stale as soon as the sensor starts making measurements.
It is possible to update the EID and replan with MPC, but this can be computationally expensive.
Because previous research uses relatively long execution and planning horizons, we focus on a single ergodic trajectory generated from an initial EID.

Submodularity seems to be a possible justification for ergodic control.
We next examine a particular type of submodularity that best justifies ergodic trajectories.

\subsection{Example and Problem Class}
\label{sec:class}

Suppose a sensor is in a domain where information is concentrated at two states.
The left state has an information density of 80\%, and the right state has a density of 20\%.
By definition, an ergodic trajectory splits its time proportionally to this ratio, and this falls out of the metric in \cref{eq:metric}.
Perfect ergodicity (i.e., $\mathcal{E} = 0$) can be achieved if $\ck=\phik$:
$$
\frac{1}{T}\sum_{\point_d\in X_d}\tau(\point_d)\Fk(\point_d) = \sum_{\point_d\in X_d} \phi(\point_d)\Fk(\point_d)\text,
$$
where $X_d$ is a discrete set of states with nonzero information, $\tau(\point_d)$ is the time spent in state $\point_d$, and $\phi(\point_d)$ represents the information at $\point_d$.
Equality holds when
$$
\frac{\tau(\point_d)}{T} = \phi(\point_d)\text.
$$
That is, perfect ergodicity is achieved if the proportion of time spent at $\point_d$ is equal to the information at that location.
In our example, the sensor spends $0.8T$ in the left state and $0.2T$ in the right.
After spending $0.8T$ at the left state, the ergodic trajectory never returns.
One situation where this behavior is optimal is if the state is stripped of information after $0.8T$.
Then, the 20\% state will contain more information after $0.8T$, and an optimal trajectory will spend the rest of the time there.


Using the above example as a guide, we claim that an ergodic trajectory minimizes the time to gather all available information in a domain if the following model for information collection and submodularity holds:
\begin{enumerate}
	\item Information is collected (and depleted) from a state when a sensor spends time there.
	\item Information is collected from all states at the same rate: $1/T$ per unit time for a continuous trajectory and $1/N$ per time step for a discrete trajectory with $N$ steps.
	\item The information available at state $\point$ is equal to $\phi(\point)$. In a discrete domain, we assume $\sum_{\point_d\in\domain_d} \phi(\point_d)=1$ (the analog to $\int_{\domain} \phi(\point)\dpoint=1$ in the continuous case).
\end{enumerate}

\subsection{Time Horizon Selection}
\label{sec:horizon_example}
Our problem class requires a linear collection (and depletion) of information.
If we know the rate at which information is collected at, we can choose the ergodic trajectory horizon to efficiently collect the available information.

Assume we have the same two-state example from the previous subsection, where the left and right states have 0.8 and 0.2 units of information, respectively.
Assume further that we know the collection rate is 0.1 per step; at each step, the sensor collects 0.1 information units from its current state.
There is a cost to switch between the states and the sensor starts in the left state.

The trajectory that minimizes the time and cost to collect all information is 10 steps long.
The sensor spends its first eight steps in the left state and its last two in the right state.
This perfectly ergodic trajectory collects all information available while minimizing the switching cost.

If we instead generated a 20-step ergodic trajectory minimizing control cost, the resulting trajectory would spend 80\% of its time in the left state and 20\% in the right---so, 16 steps in the left state followed by four in the right.
After its first eight steps in the left state, the sensor would deplete all available information there.
It would collect no new information until switching to the right state.
Eventually, all information would be collected, but it would have taken roughly twice as long as with the 10-step horizon.

If we picked a shorter horizon, like five steps, a perfectly ergodic trajectory would spend four steps in the left state and then one in the right.
However, at the end of this trajectory, the sensor would only have collected half the available information---the left state would still have 0.4 and the right would have 0.1.
The sensor could execute another five-step ergodic trajectory starting from the sensor's last position (the right state).
This trajectory would spend one step in the right state followed by four in the left.
After the two five-step trajectories, all information is collected---just as it was at the end of our single 10-step trajectory.
However, the sensor incurs twice the cost by switching states twice, using two sweeps to cover the space.
Further, two ergodic trajectories are computed instead of one, which can be expensive.

By selecting a horizon for our ergodic trajectory, we assume a decay rate.
If this rate matches the true decay rate, we can minimize the time required to collect all available information.
In many dynamical systems, a trajectory with this carefully selected horizon will also minimize the control effort required to gather all information, as it did in our example.
However, this is not the case with all dynamical systems.
For example, an oscillating system might trade time for energy use.
In these systems, an ergodic trajectory might exert extra control effort to drive the sensor to distribute measurements ergodically.

\subsection{Example Outside the Class}
\label{sec:violation}

Consider two observation posts on either side of a runway.
An observer estimates the distance to an approaching aircraft.
From either post, the observer measures the true distance corrupted with zero-mean Gaussian noise.
The left post offers the best view, while the right post is blocked by trees.
As a result, the Gaussian noise of the left post has standard deviation $\sigma_{\text{small}}$, and the noise of the right has $\sigma_{\text{large}}>\sigma_{\text{small}}$.

Both posts have non-zero information density---from either, enough noisy measurements can be stitched together to give a low variance distance estimate.
However, more observations are required from the right (noisier) post.
The optimal search trajectory makes all measurements from the left post.
However, an ergodic trajectory would spend some time in the right post because it has non-zero information density.
The linear information decay assumed in our problem class implies all information will be ``used up" from the left post after some fraction of the time horizon.
As a result, an ergodic trajectory reserves some time for the right post.

The ergodic trajectory is suboptimal because it falls outside of our problem class.
The sensor model implies measurements from the left post are \textit{always} more informative than those from the right, regardless of the time spent there.

\subsection{Analysis of the Ergodic Metric}
\label{sec:proof}
So far, we have provided intuitive arguments for the connection between submodularity and the optimality of ergodic trajectories.
In this section, we provide a theoretical argument using the Fourier-based ergodic metric.

Before proceeding, consider two preliminaries.
First, the Fourier transform is linear with respect to distributions.
That is, if $z,y\in\mathbb{R}$, and $\phi^1$ and $\phi^2$ are two distributions, then
$$
\phi = z\phi^1 + y\phi^2 \iff \phik = z\phik^1 + y\phik^2\text.
$$
Second, when adding two distributions, we add the densities at each point; scaling a distribution scales the density at each point.
When adding or scaling distributions, the resulting distributions will not integrate to 1, so care must be taken when performing these operations.

Our argument proceeds as follows.
Suppose we desire a trajectory with horizon $T = \Ta + \Tb$ that is split into two partial trajectories $\xa$ and $\xb$.
Suppose $\xa$ has already been executed for its horizon $\Ta$.
This partial trajectory has a spatial distribution $\ca$ and coefficients $\cka$, each of which are normalized by horizon $\Ta$.
We want to design the remainder of the trajectory, $\xb$, for the remaining horizon $\Tb$ so that the entire trajectory $x = \xa + \xb$ is ergodic.
The coefficients for each partial trajectory are
\begin{equation}
	\begin{aligned}
		\cka &= \frac{1}{\Ta}\int_0^{\Ta}\Fk(x(t)) \dt\text,\\
		\ckb &= \frac{1}{\Tb}\int_{\Ta}^{\Ta+\Tb}\Fk(x(t)) \dt\text.
	\end{aligned}
\end{equation}
The coefficients for the entire trajectory are a weighted average of the coefficients for the individual trajectories:
\begin{equation}
	\begin{aligned}
		\ck &= \frac{1}{\Ta + \Tb}\int_0^{\Ta + \Tb} \Fk(x(t)) \dt\\
			   &= \frac{1}{\Ta + \Ta} \left(\Ta \cka + \Tb \ckb\right)\text.
	\end{aligned}
\end{equation}
The objective function then becomes

\begin{equation}
	J(\xb) = \sum_\kk \Lambdak \left(\frac{\Ta \cka + \Tb \ckb}{\Ta + \Tb} - \phik\right)^2\text.
	\label{eq:proof1}
\end{equation}
We can reorder this objective so it becomes
\begin{equation}
	J(\xb) = \left(\frac{\Tb}{\Ta+\Tb}\right)^2 \sum_\kk \Lambdak \left(\ckb - \phik'\right)^2\text,
\end{equation}
where
\begin{equation}
	\phik' = \frac{\Ta+\Tb}{\Tb}\left(\phik - \frac{\Ta}{\Ta+\Tb} \cka\right)\text.
	\label{eq:newphik}
\end{equation}
We drop the scale factor, yielding the equivalent objective
\begin{equation}
	J(\xb) = \sum_\kk \Lambdak \left(\ckb - \phik'\right)^2\text.
	\label{eq:newproof}
\end{equation}
Therefore, designing $\xb$ to minimize \cref{eq:proof1} is equivalent to designing $\xb$ to minimize \cref{eq:newproof}.
We are effectively designing $\xb$ to be ergodic with respect to a new distribution $\phi'$, whose coefficients are $\phik'$.
Because of the linearity of the Fourier transform, the modified distribution $\phi'$ is similar to the modified coefficients $\phik'$:
\begin{equation}
	\phi' = \frac{\Ta + \Tb}{\Tb}\left(\phi - \frac{\Ta}{\Ta + \Tb}\ca \right)\text.
	\label{eq:newphi}
\end{equation}
The distribution $\phi'$ results from the effect of partial trajectory $\xa$ and its corresponding distribution $\ca$ on the original distribution $\phi$.
The quantity inside the parentheses of \cref{eq:newphi} is equal to the original distribution minus a scaled version of $\ca$; the scale factor is equal to the proportion of time spent in trajectory $\xa$.

However, the distribution in the parentheses of \cref{eq:newphi} is invalid because it does not integrate to 1.
If we are designing $\xb$ to be ergodic with respect to spatial distribution $\phi$, we normalize $\cb$ and $\phi$ so we can compare them.
The linearity of the Fourier decomposition implies
\begin{equation}
	\int_X \left(\phi(\point) - \frac{\Ta}{\Ta + \Tb} \ca(\point)\right) \dpoint = \frac{\Tb}{\Ta + \Tb}\text.
\end{equation}
Therefore, we have the normalization term $(\Ta+\Tb)/\Tb$ in \cref{eq:newphi}, ensuring $\phi'$ integrates to 1.

We have shown that the ergodic objective from \cref{eq:metric} reduces the values of states in which time has already been spent, proportional to the time spent there; this result matches the conditions presented in \cref{sec:class}.

These results satisfy an intuitive result: if $\Ta=\Tb$ and $\cka=\phik$, then $\phik' = \phik$.
That is, if the partial trajectory $\xa$ is perfectly ergodic, then $\xb$ should be ergodic with respect to the same distribution in order for the whole trajectory to be ergodic.
The trajectory $\xa$ collects half the information available at every state, so it makes sense to perform a similar sweep over the domain to retrieve the remaining information.

Consider another intuitive result.
From \cref{eq:newphi}, $\phi'(\point)<0$ if
$$
\phi(\point) < \frac{\Ta}{\Ta + \Tb}\ca(\point)\text.
$$
If this is the case, we have oversampled point $\point$ during partial trajectory $\xa$ and it is impossible to rectify this in the remaining horizon $\Tb$~\cite{ivic2016}.
It is possible to overcome this oversampling by increasing the horizon $\Tb$, which would ensure a smaller scale applied to $\ca(\point)$.

\subsection{Spatial Correlation}
\label{sec:space}
Our intuitive examples used domains where information is concentrated in a discrete set of states so we could observe the effect of sampling from a state.
This observation is more difficult in a continuous domain.
Even with noiseless dynamics, the agent cannot sample all states in a continuous domain in finite time.
In a real scenario with noise, the vehicle will likely never return to the same exact state, so the notion of spending more time in a state is unrealistic.

These problems arise from use of the Dirac delta in the definition of the time-averaged statistics $c$, which sets the sensing footprint at any time to be a single state.
An alternative is to encode a larger sensor footprint into $c$~\cite{ayvali2017}.
For example, if a sensor gathers information from all points within a radius of its current state, the time-averaged statistics $c$ can be defined to reflect this.
However, the bulk of existing work uses the Dirac delta, so we use it here.

Although the Dirac delta implies no spatial correlation between measurements, correlation is introduced by the ergodic metric, giving the sensor a footprint larger than a single state.
We have assumed a perfect relationship between a spatial distribution $\phi$ and its coefficients $\phik$---that is, decomposing $\phi$ into coefficients $\phik$ and using these coefficients to reconstruct a spatial distribution would lead to $\phi$.
This interchangeability holds as $K\to\infty$, but real implementations use a finite number of coefficients, yielding a band-limiting effect on the representational power of the Fourier decomposition~\cite{millerthesis2}.
It has been posited that this effect can be beneficial as it allows for unmodeled uncertainty in the EID.
We build on this idea, suggesting that fewer coefficients can add spatial correlation between vehicle states, as higher-order coefficients are needed to capture fine differences in a distribution or trajectory.

\begin{figure}
	\centering
	\includegraphics[width=1.6in,trim={1.1in 0.2in 1.1in 0.3in},clip]{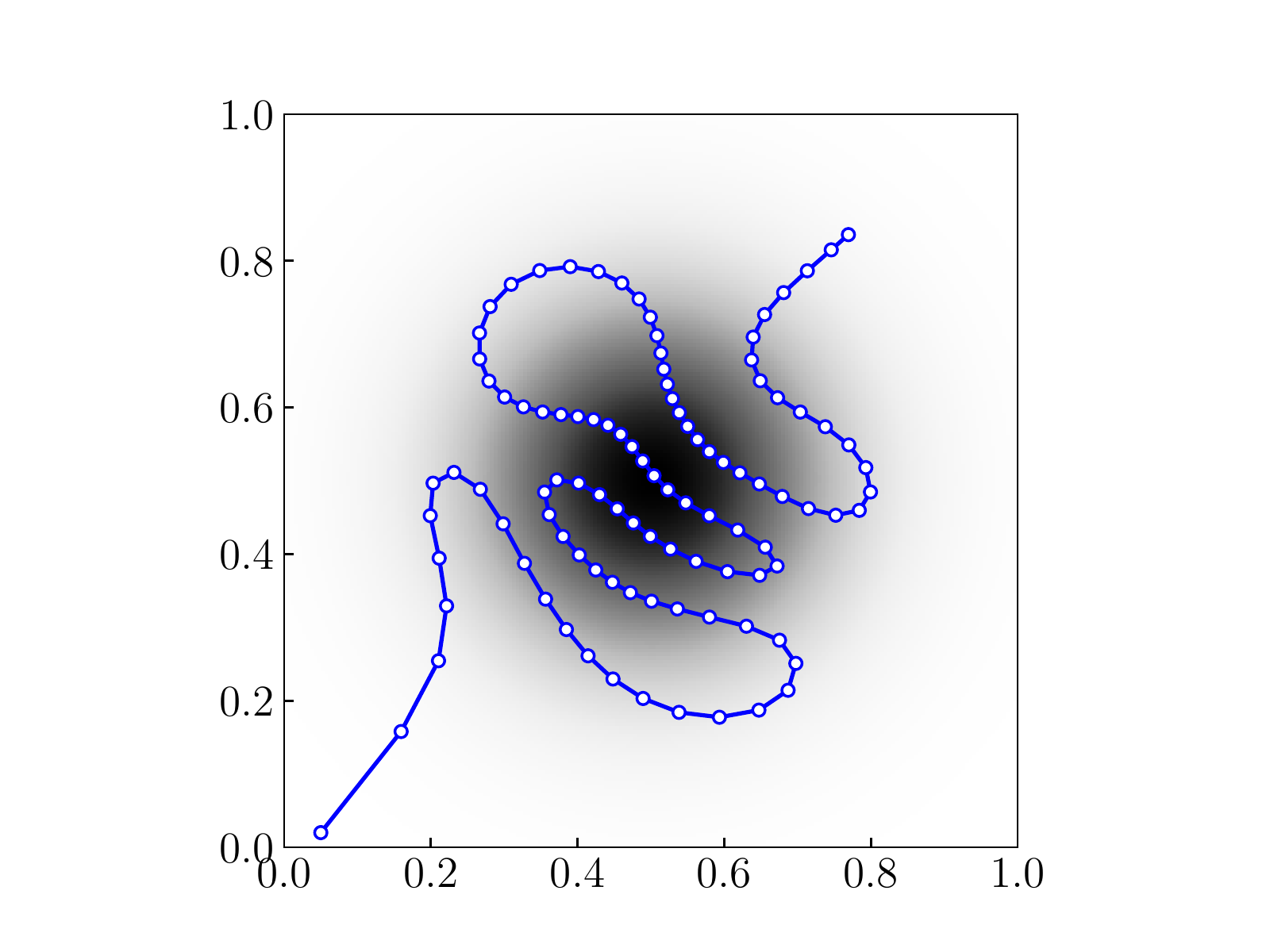}
	\includegraphics[width=1.6in,trim={1.1in 0.2in 1.1in 0.3in},clip]{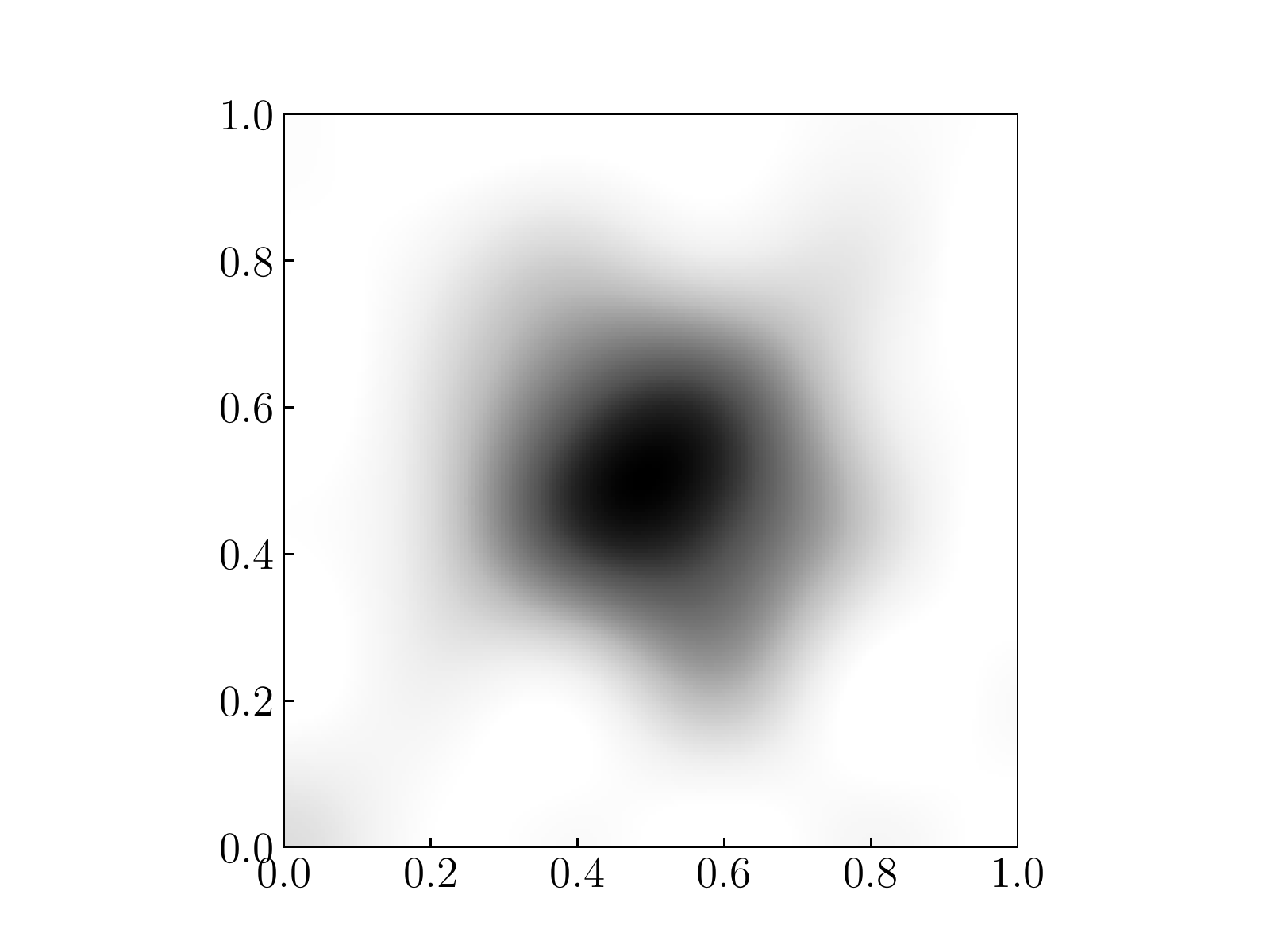}
	\includegraphics[width=1.6in,trim={1.1in 0.2in 1.1in 0.3in},clip]{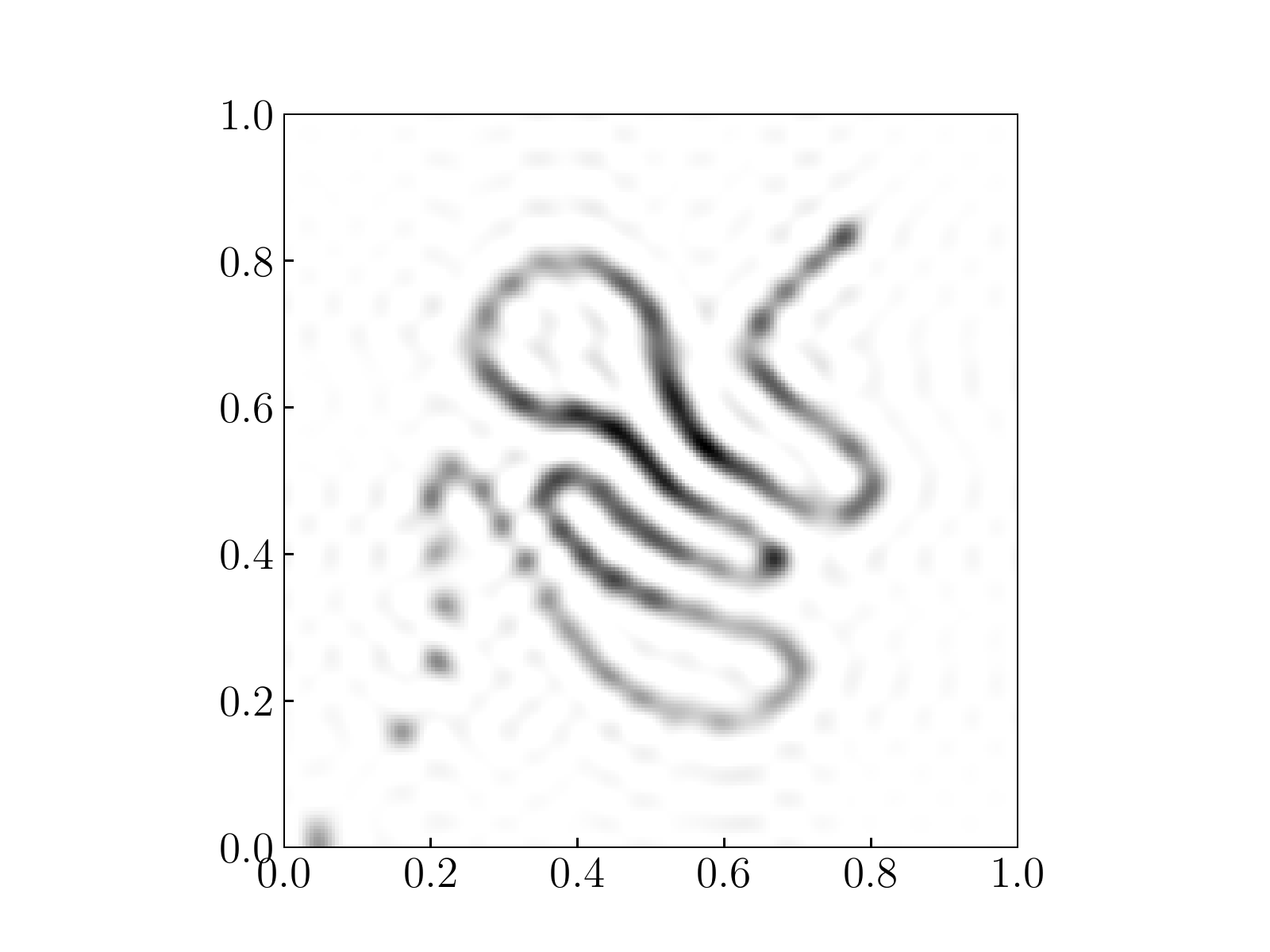}
	\includegraphics[width=1.6in,trim={1.1in 0.2in 1.1in 0.3in},clip]{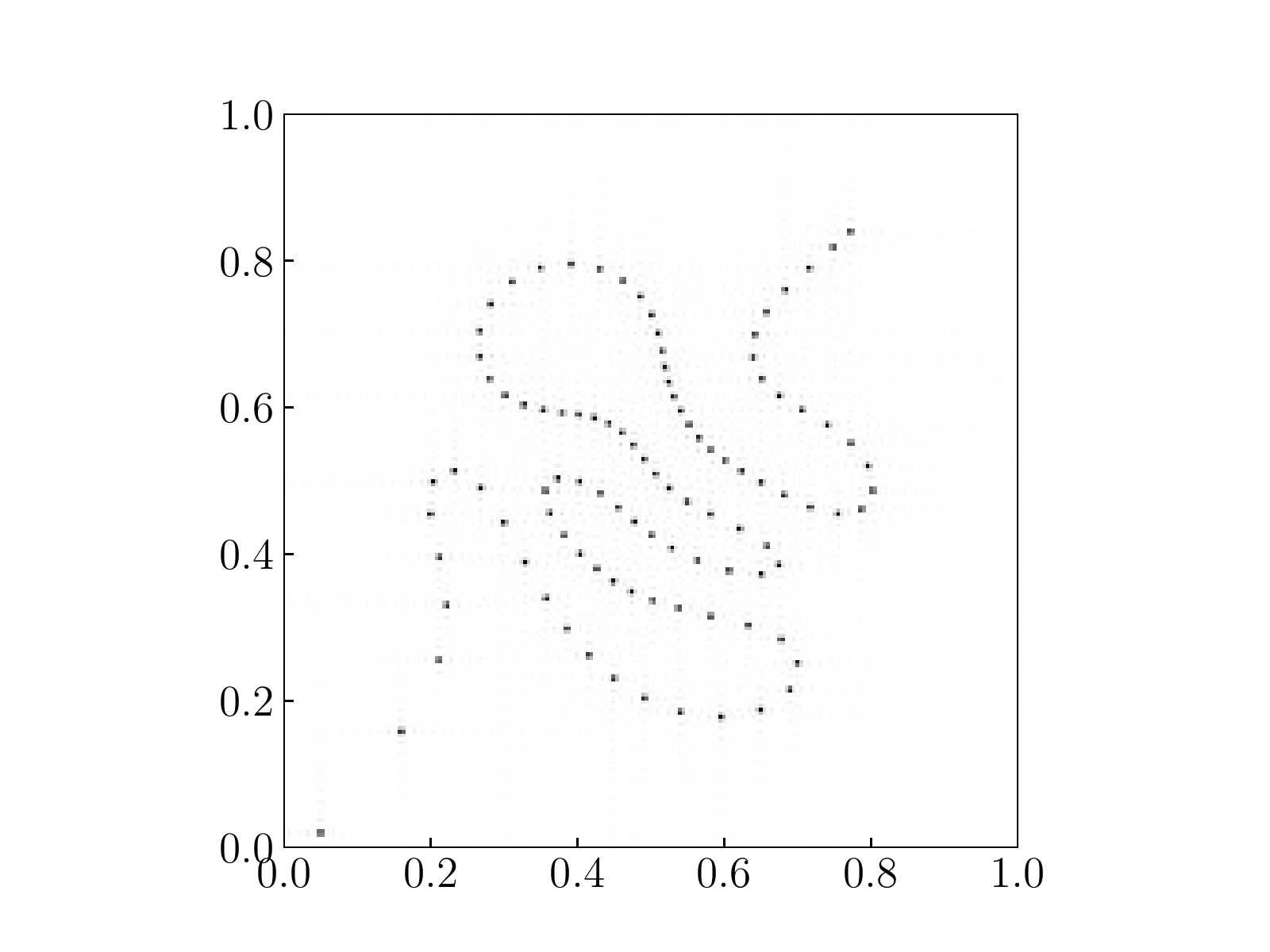}
	\caption{In the upper left, the original distribution and a trajectory designed to be ergodic with respect to it. The reconstructed distributions from this trajectory when using $K=5$, $K=30$, and $K=150$ coefficients are shown in the upper right, lower left, and lower right, respectively.}
	\label{fig:keffect1}
\end{figure}

An example of this spatial correlation is shown in \Cref{fig:keffect1}.
A discrete ergodic trajectory is generated for a simple Gaussian distribution.
This trajectory is decomposed into sets of coefficients $\ck$ for different numbers of coefficients $K$.
These sets of coefficients are used to reconstruct spatial distributions of the trajectory.
When $K=5$, the resulting spatial distribution of the trajectory looks fairly similar to the original Gaussian distribution.
When $K=30$, the spatial distribution more closely matches the trajectory.
When $K=150$, the spatial distribution is so similar to the trajectory that individual points along the trajectory are discernible.
Visually, the coarse $K=5$ distribution most closely matches the original spatial distribution.
Even though a small number of states are visited in the trajectory, much of the state space has positive density because of the spatial correlation introduced by the small number of coefficients.
In contrast, there is much less spatial correlation in the $K=150$ distribution; only states in the near vicinity of the discrete trajectory's points have any density.

\begin{figure}
	\centering
	\includegraphics[width=1.6in,trim={1.1in 0.2in 1.1in 0.3in},clip]{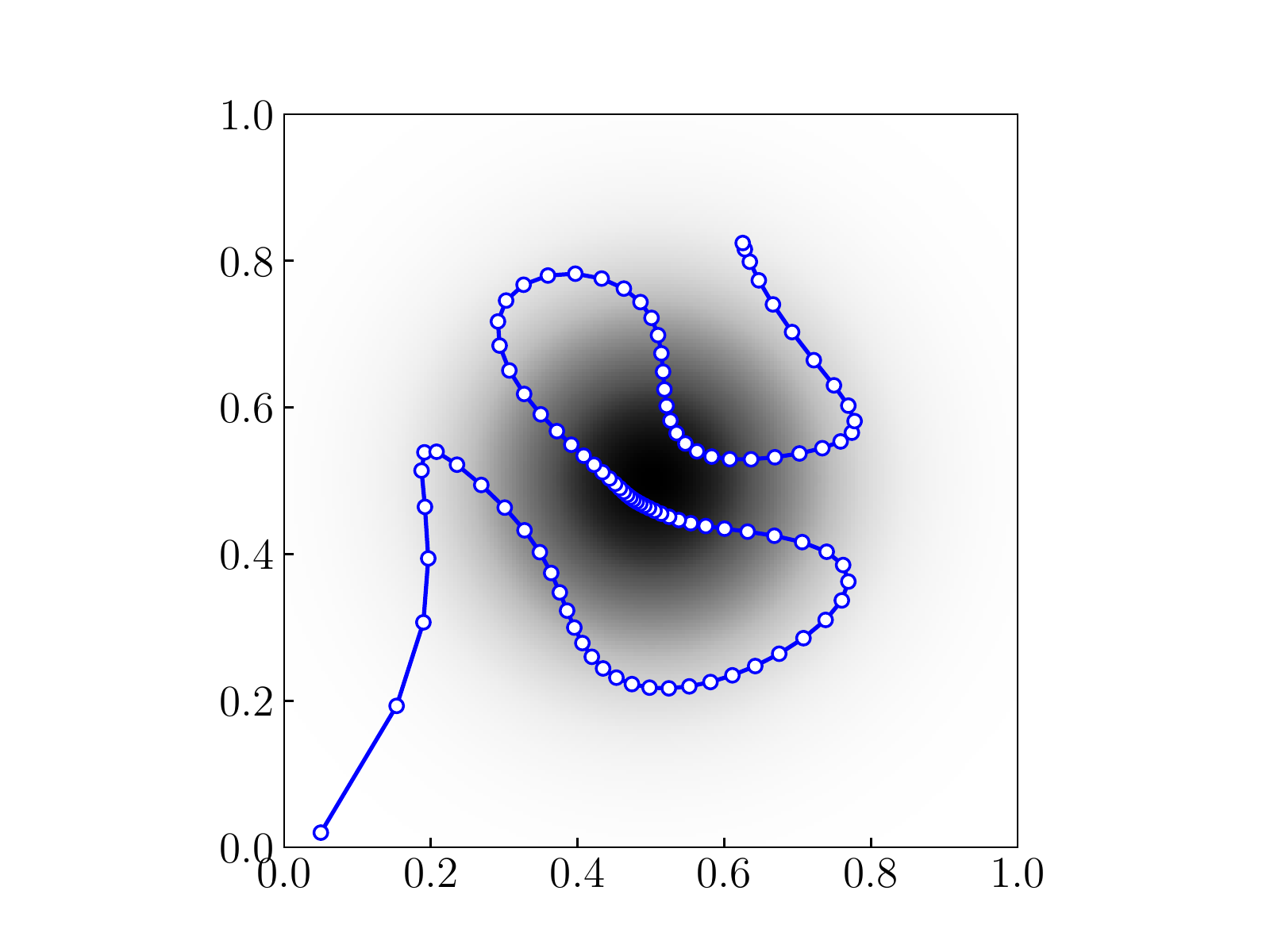}
	\includegraphics[width=1.6in,trim={1.1in 0.2in 1.1in 0.3in},clip]{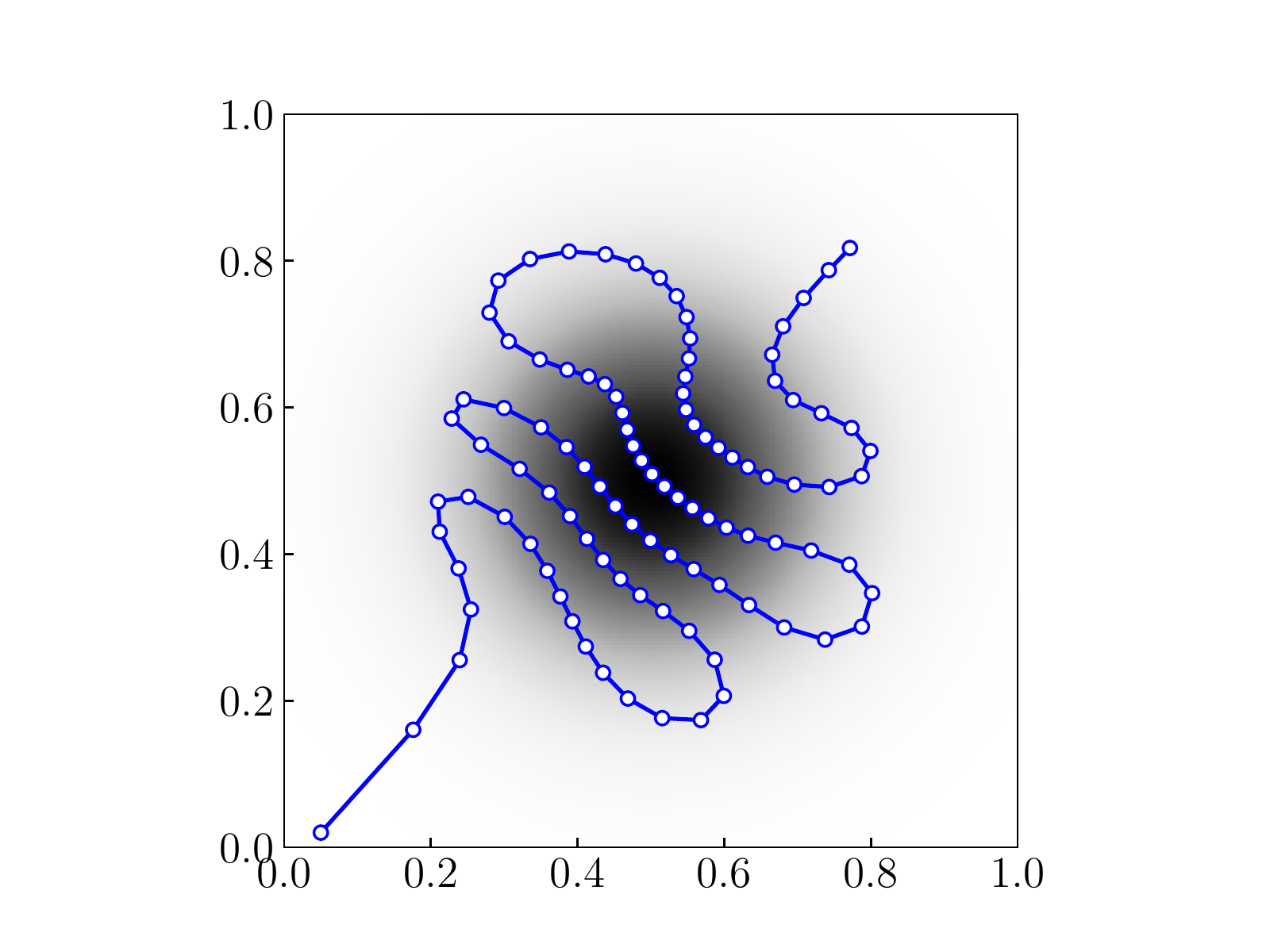}
	\caption{Trajectories generated to be ergodic with respect to a Gaussian distribution. The left trajectory was generated with $K=5$ coefficients, and the right was generated with $K=100$.}
	\label{fig:keffect2}
\end{figure}

This spatial correlation affects the ergodic trajectories generated.
\Cref{fig:keffect2} shows trajectories generated for the same distribution $\phi$, but one uses $K=5$ coefficients and the other uses $K=100$.
The trajectories are generated using PTO until a descent direction threshold is reached~\cite{miller2013}.
The trajectory generated with fewer coefficients is more spread out, because the coarse decomposition implies greater spatial correlation.

\begin{figure}
	\centering
	\includegraphics[width=1.6in,trim={1.1in 0.2in 1.1in 0.3in},clip]{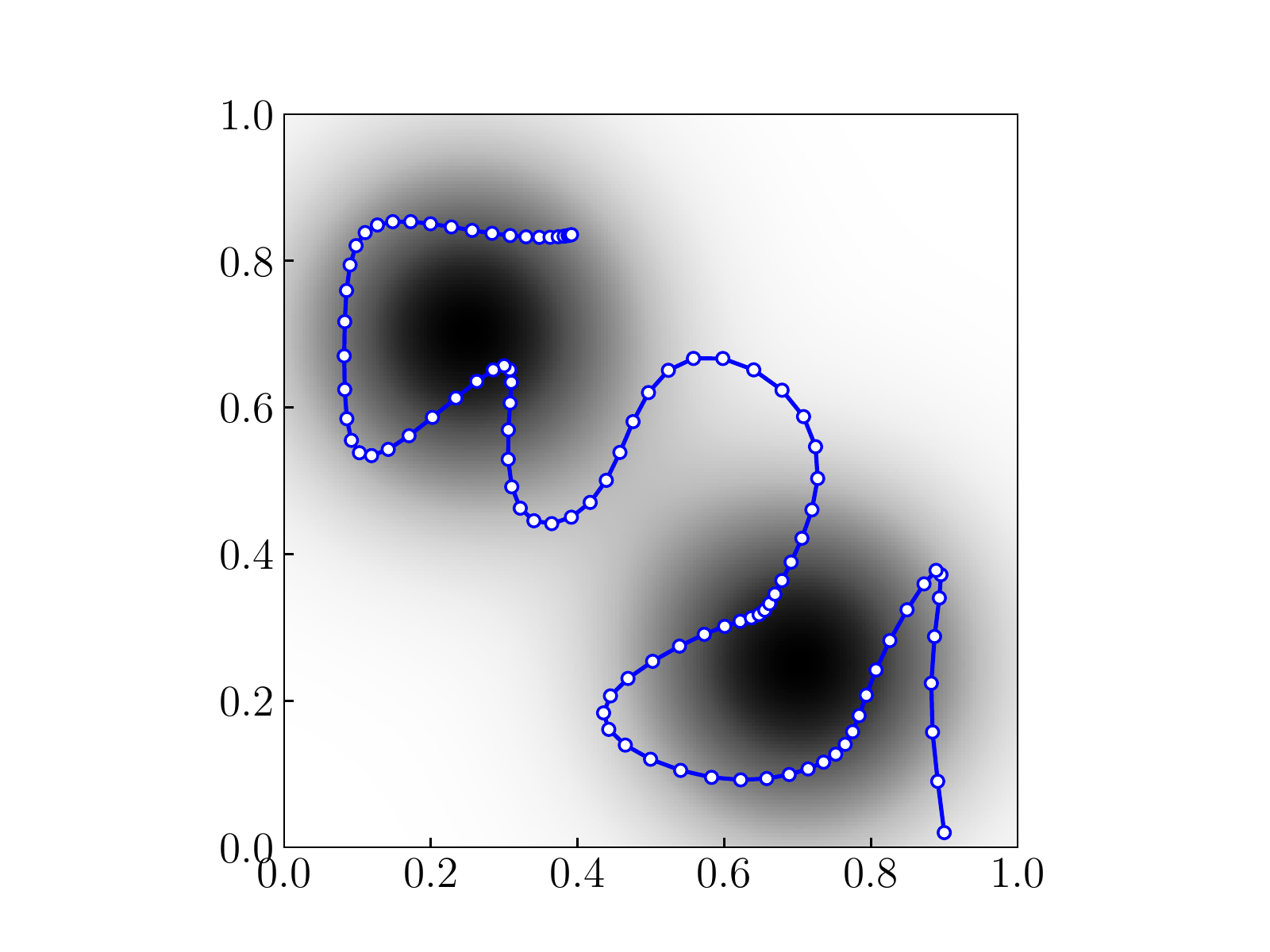}
	\includegraphics[width=1.6in,trim={1.1in 0.2in 1.1in 0.3in},clip]{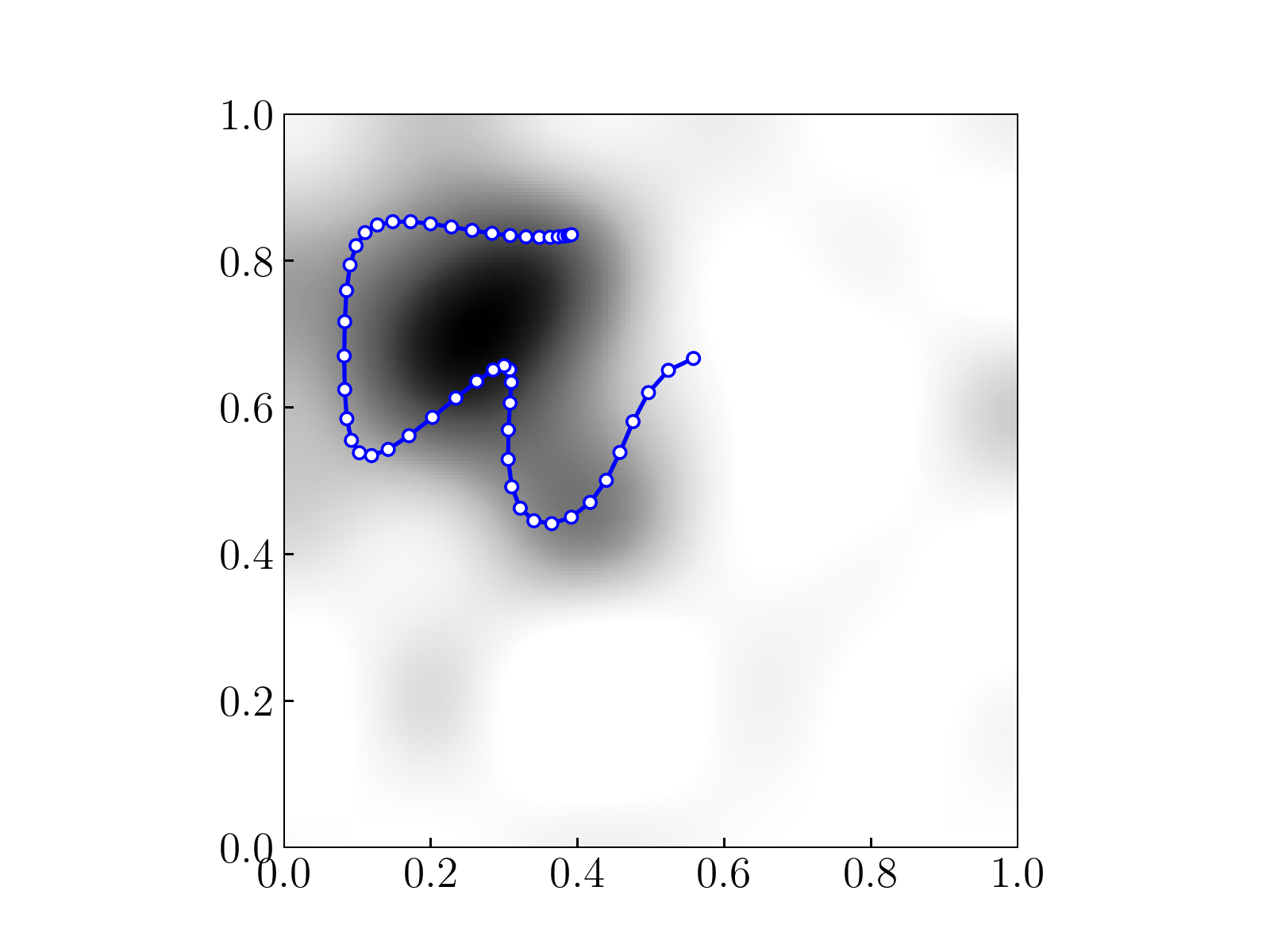}
	\caption{On the left, a trajectory ergodic with respect to a bimodal distribution $\phi$ starts in the lower right corner. On the right, we show the modified spatial distribution according to \cref{eq:newphi} after half the trajectory is executed. The lower right mode is gone because all information was collected after the first half of the trajectory was spent there.}
	\label{fig:keffect3}
\end{figure}

\Cref{fig:keffect3} shows an example of the partial-trajectory example from the previous subsection.
Although the first partial trajectory only coarsely covers the lower-right mode, the modified spatial distribution suggests all information was gathered from the mode.

\section{INFORMATION GATHERING EXPERIMENTS}
\label{sec:examples}
We use two experiments to test the relationship between ergodicity, submodularity, and information gathering.
In each experiment, ergodic trajectories are generated for a mobile sensor and an EID composed of one or two Gaussians.
The EID covers the unit square, which is discretized into a $10\times 10$ grid.
The information in each cell is obtained from the EID.

At each time step, the sensor collects (and removes) information from the cell it occupies at a specified rate.
If there is not enough information in the cell, the sensor collects whatever is left.
Discretization implies spatial correlation between measurements, as measurements from any point in a cell affect future measurements from any point inside the same cell.
Note that this spatial correlation does not correspond to the spatial correlation implied by a finite number of coefficients.
This domain is simple but not unlike others that have been used in information-gathering research~\cite{hollinger2014}.

We use PTO with $K=50$ to generate discrete trajectories with $N=100$ points starting from $(0.25,0.35)$.
Single integrator dynamics are used with a time step of 0.5 seconds.

\subsection{Ergodic Score and Information Collected}

We have claimed that perfectly ergodic trajectories are information-optimal under linear information submodularity.
If our claim is correct, information gathered should increase as ergodic score improves (i.e., $\mathcal{E}$ decreases).

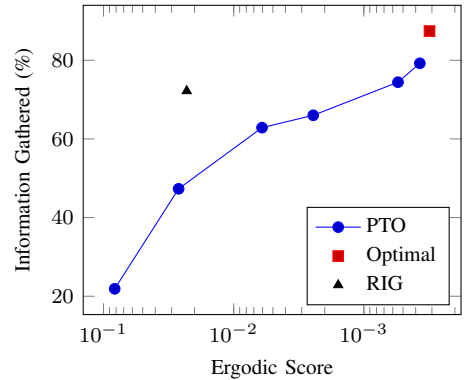
\begin{figure}
	\centering
    \begin{tikzpicture}[font=\footnotesize]
        \begin{axis}[legend pos = {south east}, ylabel = {Information Gathered (\%)}, xlabel = {Ergodic Score}, x dir=reverse, xmode = {log}, width = {2.6in}]\addplot+ coordinates {
    (0.08204990592667792, 21.889518334657133)
    (0.026489327476043677, 47.31695185832849)
    (0.006055783734311496, 62.85940787569562)
    (0.0024476719967236426, 66.00707790670519)
    (0.0005468018919614195, 74.42409681676398)
    (0.00037152296934595073, 79.22359323808918)
    };
    \addlegendentry{PTO}
        \addplot+ [only marks]coordinates {
    (0.00031294852145664903, 87.41762562092174)
    };
    \addlegendentry{Optimal}
        \addplot+ [mark={triangle*},black,mark options={black},only marks]coordinates {
    (0.02295, 72.19)
    };
    \addlegendentry{RIG}
    \end{axis}

    \end{tikzpicture}
	\caption{Information gathered as a function of ergodic score.}
	\label{fig:scoreplot}
\end{figure}

\begin{figure}
	\centering
    \includegraphics[width=1.6in,trim={1.1in 0.2in 1.1in 0.3in},clip]{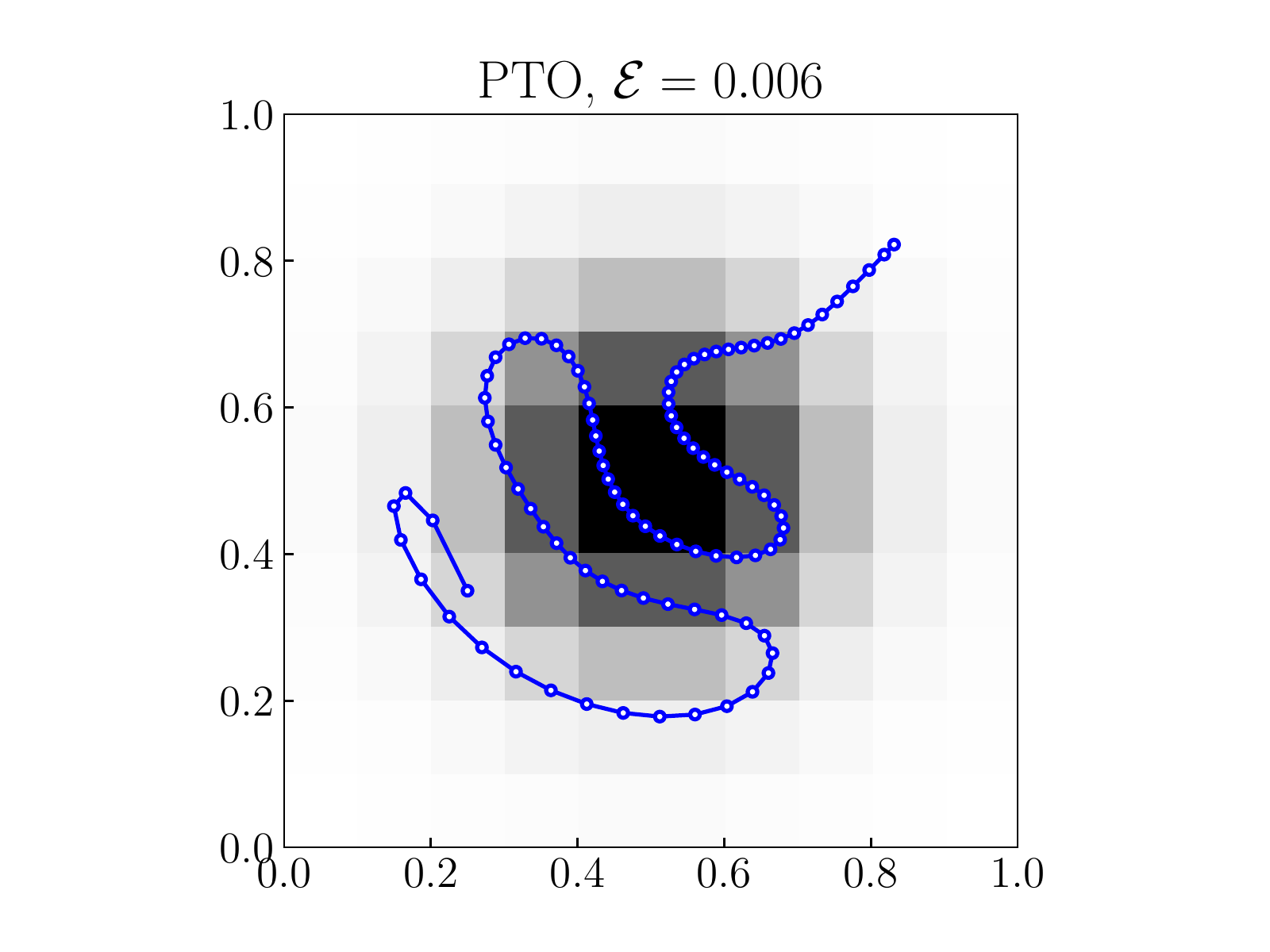}
	\includegraphics[width=1.6in,trim={1.1in 0.2in 1.1in 0.3in},clip]{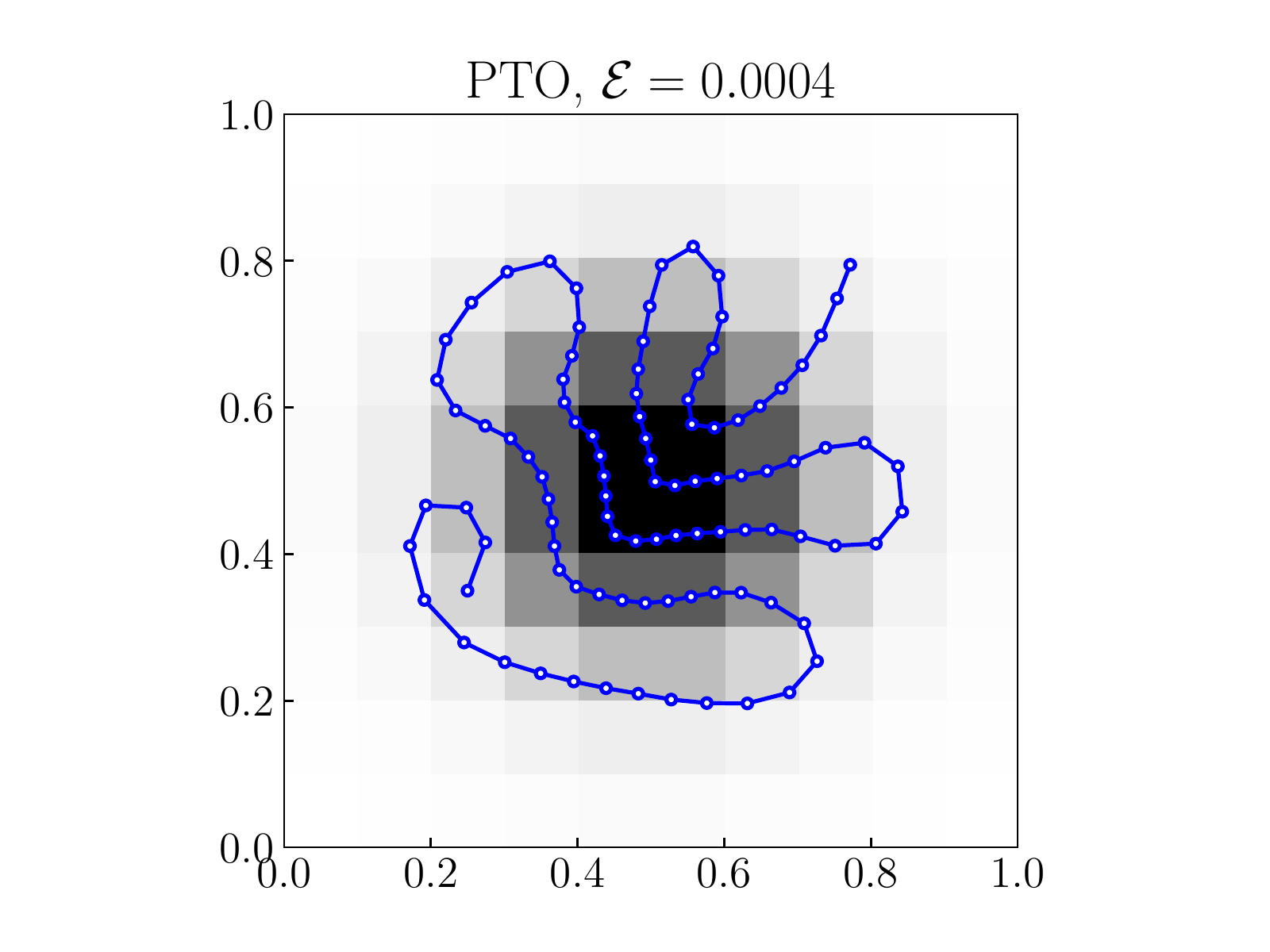}
	\includegraphics[width=1.6in,trim={1.1in 0.2in 1.1in 0.3in},clip]{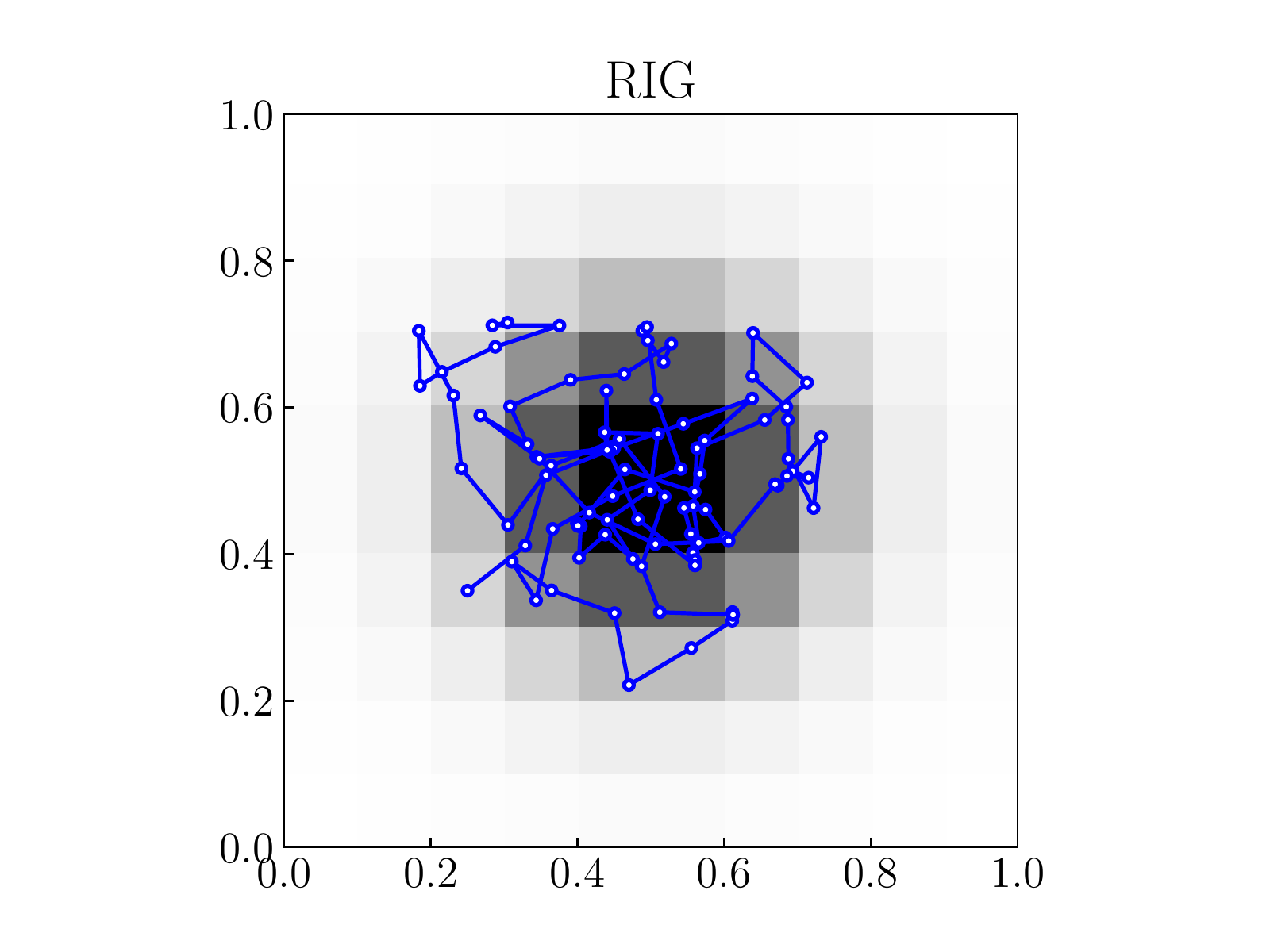}
	\includegraphics[width=1.6in,trim={1.1in 0.2in 1.1in 0.3in},clip]{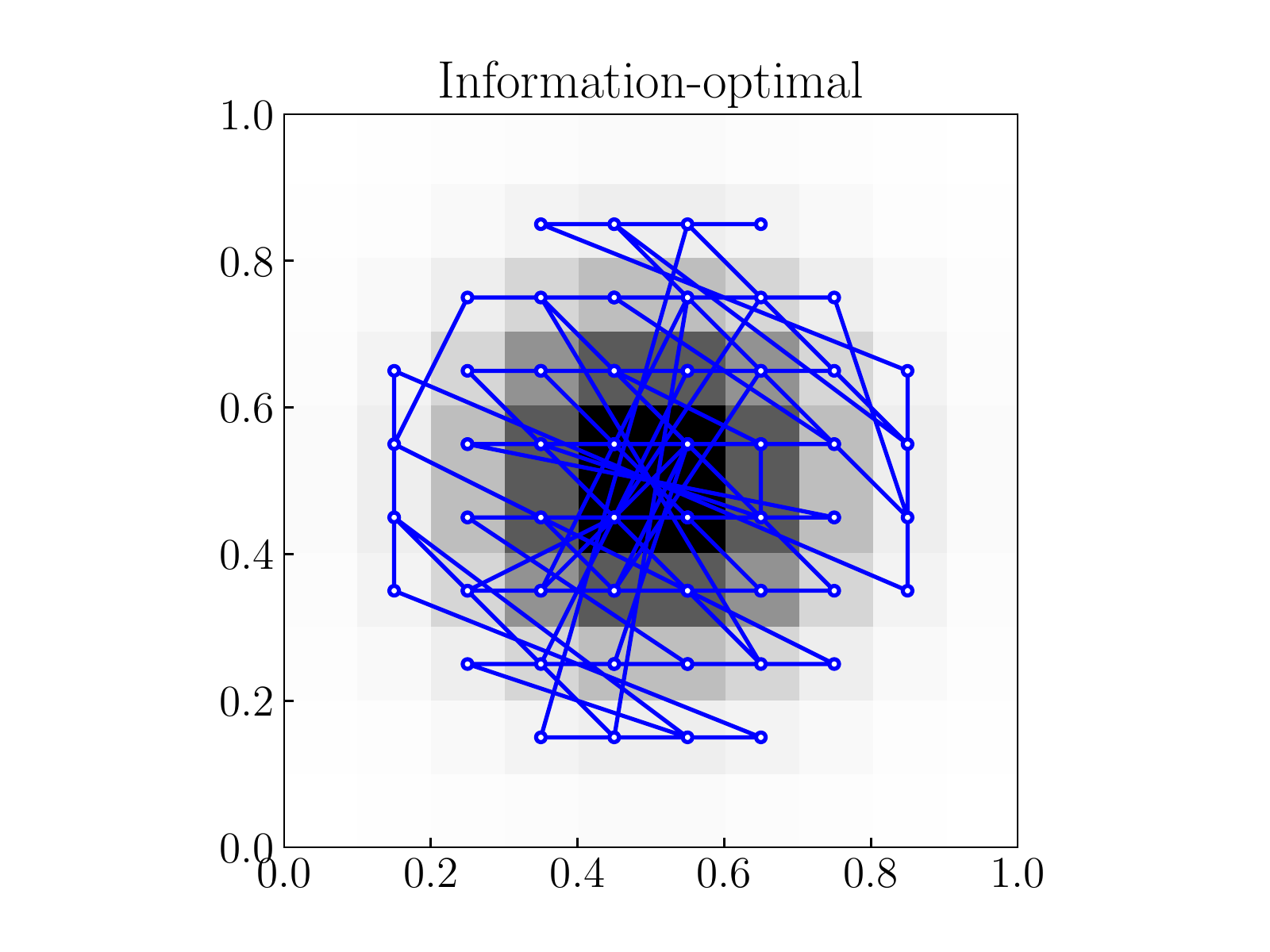}
	\caption{Trajectories generated with different methods collecting information in a discrete $10\times 10$ grid.}
	\label{fig:scoreexamples}
\end{figure}

We terminate PTO at different ergodic scores and record the information collected by each trajectory.
We compare against rapdily-exploring information gathering (RIG), a sampling-based motion planner that incorporates information submodularity~\cite{hollinger2014}.
We also generate an information-optimal trajectory that moves the sensor to the grid cell with the most information left.
Information is collected from the cell, and the process repeats.
The resulting trajectories are feasible only because of the simple dynamics; such a technique is not applicable to general systems.
\Cref{fig:scoreplot} shows the relationship between ergodic score and information collected. Trajectories are shown in \cref{fig:scoreexamples}.

As trajectory ergodicity increases, the information collected increases.
The information-optimal trajectory collects the most information and has the lowest ergodic score, reinforcing the tie between ergodicity and information collection under our model.
It is impossible to collect all the information with the finite number of discrete steps.
However, if $N\to\infty$, the information-optimal trajectory approaches $100\%$ information gathered and $\mathcal{E}\to 0$.
RIG approaches optimality if sufficient points are used, so it performs well.

Both the information-optimal and RIG trajectories are generated with an exact model of the spatial correlation involved.
In contrast, the PTO trajectories only have a sense of the spatial correlation through the finite number of coefficients used.
However, these trajectories are still competitive if solved to a low enough ergodic score.
This result validates the claim that improved ergodicity leads to more information collected in problems with our model.

\subsection{Trajectory Horizon and Information Collected}

\begin{figure}
	\centering
	\includegraphics[width=1.6in,trim={1.1in 0.2in 1.1in 0.3in},clip]{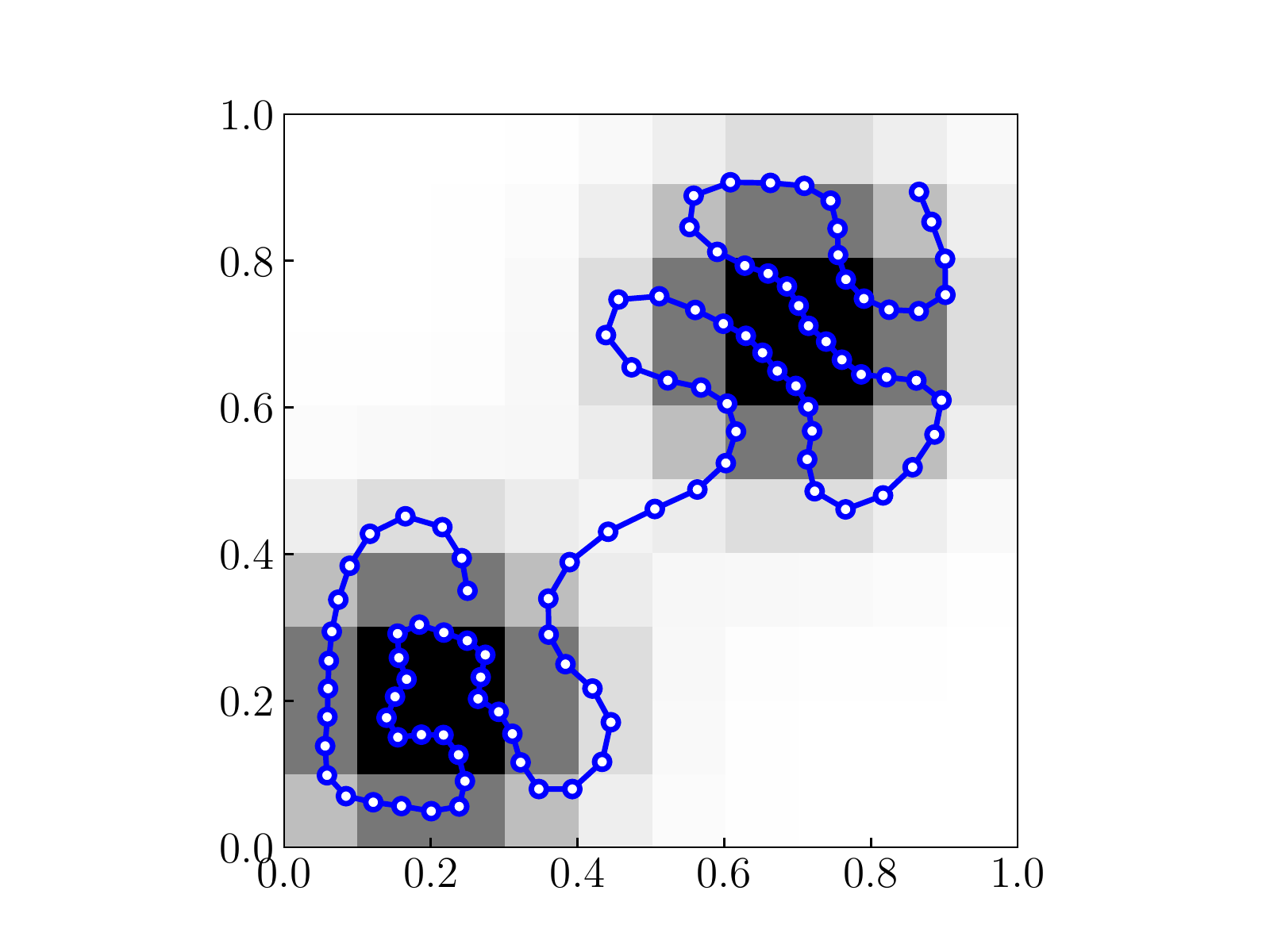}
	\includegraphics[width=1.6in,trim={1.1in 0.2in 1.1in 0.3in},clip]{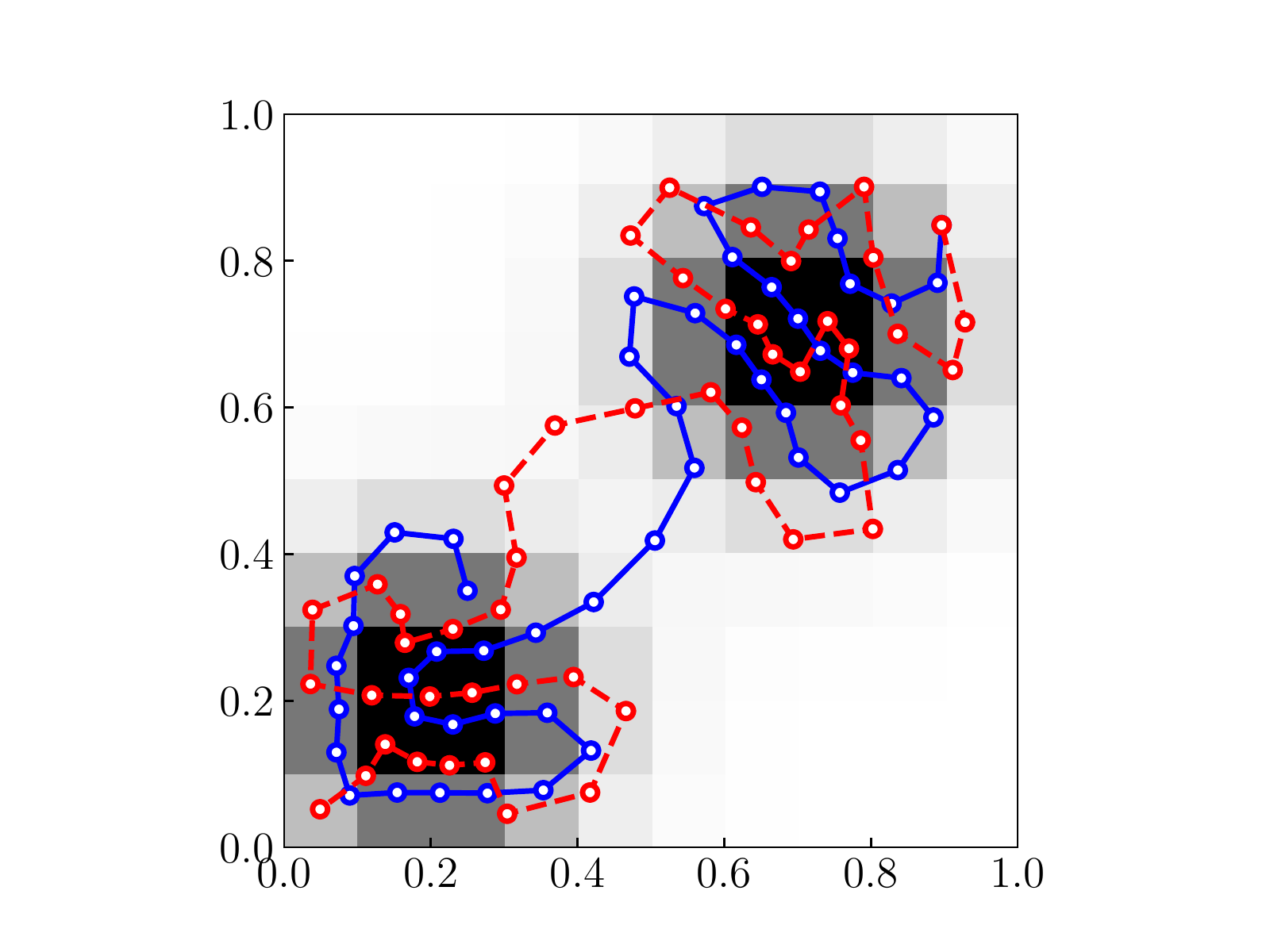}
	\caption{PTO ergodic trajectories. On the left, a single trajectory generated for horizon $N_f$. On the right, a trajectory of horizon $N_f$ is composed of two trajectories each designed for a horizon of $N_f/2$. The first sub-trajectory is the solid, blue line. The second is the red, dashed line. The single trajectory on the left collects roughly the same information with about half the cost.}
	\label{fig:double}
\end{figure}

In \Cref{sec:horizon_example}, we claim that knowledge of the information collection (decay) rate informs selection of the trajectory horizon.
Suppose we know the information collection rate is $1/N_f$ per time step, where $N_f$ is a positive integer.
When selecting a horizon $N$ for our ergodic trajectory, we posit that $N$ should match $N_f$ for the most efficient trajectories.
We set $N_f = 100$ and use PTO to generate a trajectory with $N=100$.
We then generate a composite trajectory consisting of two smaller ergodic trajectories, each with $N=50$.
\Cref{fig:double} shows the trajectories.

Both trajectories collect roughly the same information: $77.8\%$ for the single trajectory and $82.6\%$ for the composite.
However, the single trajectory has a control effort $(\sum_{n=1}^N u_n)$ of 3.6, and the composite has a control effort of 6.8.
When the horizon is too short, not enough information is collected and a second pass is needed, increasing the cost.
Thus, knowledge of the information collection rate can inform selection of the trajectory horizon.

\section{DISCUSSION}
\label{sec:conclusion}
Our work suggests ergodic control is the optimal information gathering strategy under a specific model of information collection and submodularity.
This model is unrealistic in many sensing tasks for two reasons.
First, a measurement made in one location often decreases the information available at other locations.
This spatial correlation is not captured by the traditional Dirac delta formulation in ergodic control; the spatial correlation resulting from a finite number of Fourier coefficients is unlikely to match the correlation of a real sensor.
Second, our model assumes that while some states might have more total information, information is collected at the same rate from all states.
However, in real scenarios, measurements from certain states are more informative than others.
Further, the EID is often formulated to represent the value of a single measurement, such as mutual information or the expectation of Fisher information.
Because our proposed model is unrealistic in many tasks, it is likely that ergodic control is suboptimal for general information gathering tasks; although it is possible that other models are optimally solved by ergodic control.

Ergodic control does excel in some tasks, particularly when our model of submodularity holds.
Coverage problems in which an agent must surveil or cover an area uniformly are well served by ergodic control~\cite{mathew2011}.
Coverage problems actually adhere to our problem class; there is value to visiting uncovered states, but that value decreases linearly until these states have been as well covered as others.
Ergodic control is also suitable for autonomous painting~\cite{prabhakar2016}, because painting an image is effectively a coverage problem in which the time to spend at a state is governed by the image's darkness at that state.
Because ergodic trajectories are distributed over a spatial distribution (rather than seeking out maxima or minima), they are robust to unmodeled sensor noise~\cite{millerthesis2}.

\addtolength{\textheight}{-12cm}   







\bibliographystyle{IEEEtran}
\bibliography{bib}

\end{document}